\documentclass[a4paper]{article}
\usepackage{geometry}
\usepackage{siunitx}
\usepackage{amsmath}
\usepackage{amssymb}
\usepackage{graphicx}
\usepackage{tikz}

\usepackage{authblk}
\usepackage{hyperref}

\begin{document}

\title{Precision optomechanics for geometrically modematched Fabry-Perot cavities in ALPS II}


\newcommand{\authormark}[1]{\textsuperscript{#1}}

\author{Li-Wei Wei,\authormark{1*} Kanioar Karan,\authormark{1} and Benno Willke\authormark{1}}

\maketitle

{\authormark{1}Institut f\"{u}r Gravitationsphysik, Leibniz Universit\"{a}t Hannover and Max Planck Institute for Gravitational Physics (Albert Einstein Institute), Callinstrasse 38, 30167 Hannover, Germany}

{\authormark{*}li-wei.wei@aei.mpg.de} 



\begin{abstract}
Challenging optical resonant gain achieved via the use of two geometrically modematched Fabry-Perot cavities is anticipated in Any Light Particle Search II (ALPS II) to extend the search sensitivity of light-weight sub-eV dark matter candidates to uncharted parameter space of scientific interest. We report on the precision optomechanics developped for geometrical modematching of optical cavities. The general concept, the mechanical mounting scheme, and the use of an autocollimator to streamline the metrology procedure are rigorously examined using optical interferometry, all of which meet the requirements of ALPS II.
\end{abstract}

\section{Introduction}
\begin{figure}[h!]
\centering\includegraphics[width=13cm]{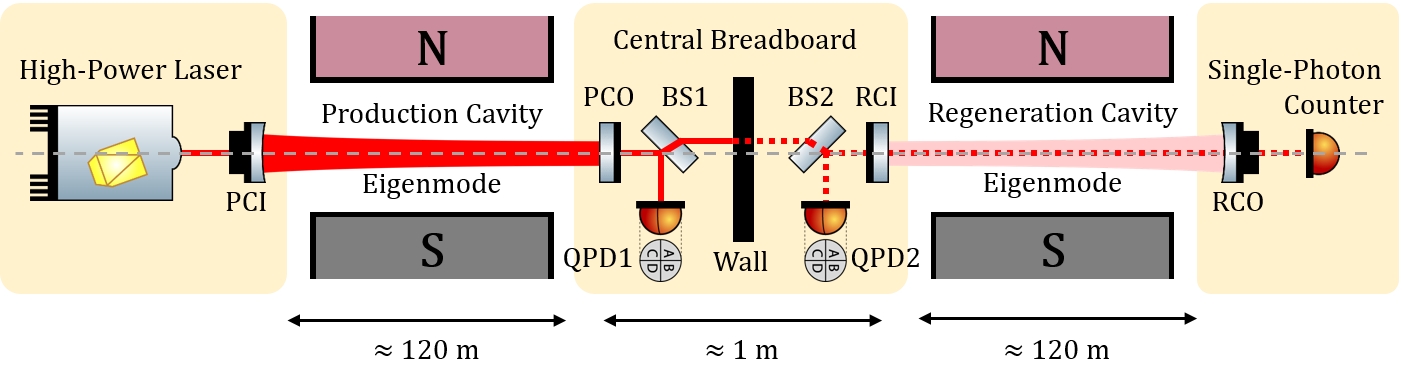}
\caption{Geometrically modematched Fabry-Perot cavities in ALPS II. In the presence of a magnetic field (symbolized by NS in the drawing), photons experience probability to be converted into ALPs on one side of the wall, and some of which can be converted back to photons on the other side of the wall. Light-shining-though-a-wall represents a photon-ALP-photon process that can be detected by a single-photon counter. Two resonant Fabry-Perot (ALP production and photon regeneration) cavities, referred to as the PC and the RC, are to be implemented to increase the event rate of such processes. As ALPs do not refract, the two cavities have to be geometrically modematched, that is, their eigenmodes should be as spatiotemporally degenerate as possible, with (red and dotted red lines) and without (dashed gray line) the presence of the modematching and beam-splitting optics. Spatial degeneracy (i.e. modematching) will be imposed by the central optical bench and is the subject of this study. PCI and RCO are curved mirrors mounted on PZT stages; PCO and RCI are planar mirrors; BS: beam-splitter; QPD: quadrant photodetector.}
\label{fig:supra_fp}
\end{figure}

Any Light Particle Search (ALPS) is a series of laboratory light-shining-through-a-wall experiments seeking light-weight sub-eV dark matter candidates including the well-motivated axion-like-particles (ALPs) \cite{Ehret+2010, Baehre+2013}. To extend the experiment sensitivity into parameter space of scientific interest in terms of ALP-photon coupling ($g_{a \gamma \gamma}$ of $\SI{2E-11}{GeV^{-1}}$, \cite{Baehre+2013, doi:10.1080/00107514.2011.563516}), in addition to using straightened dipole magnets from HERA (Hardron-Electron Ring Accelerator) at DESY Hamburg, ALPS II anticipates aggressive resonant optical gain ($\approx \num{2E8}$) via the use of two geometrically modematched Fabry-Perot cavities \cite{Baehre+2013, HOOGEVEEN19913, PhysRevLett.98.172002}, as depicted in Fig. \ref{fig:supra_fp}.

Referring to Fig. \ref{fig:supra_fp}, we apply the following concept to achieve geometric modematching of the PC and the RC:
\begin{itemize}
\item{Use plane-parallel mirror substrates for PCO and RCI and mount them parallel on a common central optical bench by means of stable precision mounting techniques. This ensures the parallelism between the optical axes of the PC and the RC eigenmodes.}
\item{Stably mount plane-parallel pick-off beam splitters on the central optical bench that steer witness beams that represent the eigenmodes of the PC and the RC onto quadrant photodetectors.}
\item{Manually align the PC eigenmode (by means of PCI) for mimimal clipping due to the finite aperture of the magnet. Then, in the absence of the wall, align the RC eigenmode (by means of RCO) such that the overlap of the two eigenmodes is maximized.}
\item{Mark the positions of the witness beams in this state on the QPDs and control PCI and RCO to keep the beams at these positions with the wall in place.}
\end{itemize}
The purpose of this study is to develop precision optomechanics that can realize such geometric modematching concept with the eventual central optical bench of ALPS II. In the next Section, we begin by deriving the tolerances on the parallelism of the optics and on the readout error of the QPDs. In Section 3 we show how parallelism of and between the plane-parallel optics can be measured and achieved with the use of an autocollimator. In Section 4 we present the results on a cavity-based eigenmode spatial overlap experiment to confirm the autocollimator measurements on the parallelism of the optics. In Section 5 we present the design and experimental results of the QPD assembly. In Section 6 we draw some conclusions and give an outlook with a stress on ALPS II implementation.

\section{Geometrical cavity modematching in ALPS II: tolerance analysis}

ALPS II anticipates by design an intra-cavity power build-up factor of \num{5000} in the Production Cavity and \num{40000} in the Regeneration Cavity, with a combined fill factor operational goal of \SI{95}{\percent} \cite{Baehre+2013, Spector:16, Pold+Spector_2017}. The goal of the ALPS II central optical bench is to constrain the angular and the lateral degrees of freedom (DoFs) to $\leq \SI{1}{\percent}$ loss in fill factor each.

On a perturbative basis about the fundamental Gaussian mode field distribution $U_0$, any angular or lateral misalignment can be regarded as power coupling to the first-order modes $U_1$:
\begin{equation}
\frac{|U_1|^2}{|U_0|^2} \approx {\left(\frac{\delta\alpha_\textrm{eig}}{\theta_{0,\textrm{eig}}}\right)}^2
 + {\left(\frac{\Delta_{\textrm{eig}}}{w_{0,\textrm{eig}}}\right)}^2,
\label{eqn:U1}
\end{equation}
where $\theta_{0,\textrm{eig}}$ is the half divergence angle and $w_{0,\textrm{eig}}$ is the cavity waist radius of the eigenmodes, and $\delta\alpha_{\textrm{eig}}$ is the angle and $\Delta_{\textrm{eig}}$ is the lateral offset between the two cavity eigenmodes; similarly, mismatches in the size or the location of the beam waists couple to the second-order modes $U_2$:
\begin{equation}
\label{eqn:U2}
\frac{|U_2|^2}{|U_0|^2} \approx {\left(\frac{\delta z_{0,\textrm{eig}}}{2 \cdot z_R}\right)}^2
 + {\left(\frac{\delta w_{0,\textrm{eig}}}{{w_{0,\textrm{eig}}}}\right)}^2,
\end{equation}
where $\delta z_{0,\textrm{eig}}$ is the difference in the waist locations and $\delta w_{0,\textrm{eig}}$ is the difference in the waist radii, and $z_R$ is the Rayleigh range \cite{Anderson_1984, Kwee+2007}.

For a nominal eigenmode waist radius of $w_{0,\textrm{eig}} \approx \SI{6.4}{\milli\meter}$ and a nominal half divergence angle of $\theta_{0,\textrm{eig}} \approx \SI{52}{\micro\radian}$ ($z_R \approx \SI{120}{\meter}$, see Fig. \ref{fig:supra_fp}), with Eqn. \ref{eqn:U1}, the $\leq \SI{1}{\percent}$ per DoF loss goal equates to $\lesssim \SI{0.64}{\milli\meter}$ in relative lateral translation and $\lesssim \SI{5.3}{\micro\radian}$ in relative angle between the two cavity eigenmodes, i.e., 
\begin{equation}
\delta\alpha_\textrm{eig} \lesssim \SI{5.3}{\micro\radian} \hspace{0.5cm} \textrm{and} \hspace{0.5cm} \Delta_\textrm{eig} \lesssim \SI{640}{\micro\meter}.
\label{eqn:ALPS_req}
\end{equation}
As one can see, due to the large waist size the requirement is much more stringent on the angular DoF than on the lateral DoF. We will refer to these as our angular/lateral specifications of the central optical bench in the following.

\section{Parallelism of and between the plane-parallel optics: mounting mechanisms and autocollimator measurements}

\begin{figure}[h!]
\centering
\includegraphics[scale=0.28]{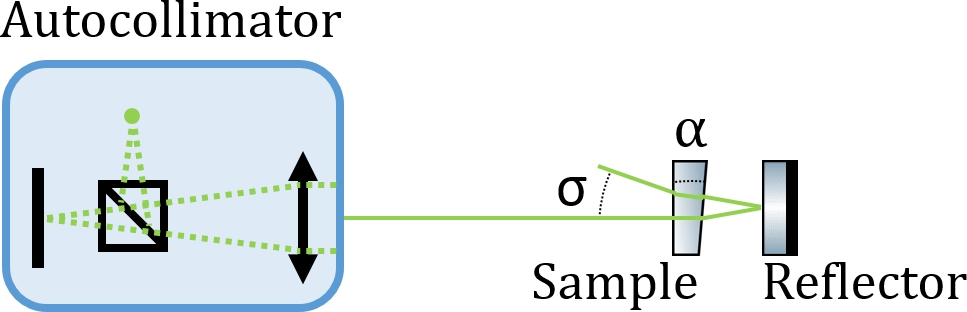}
\caption{Double-pass wedge angle measurement of sample optics. The planar high reflector is first adjusted to arrive at normal incidence. The transparent sample is then inserted. The wedge angle $\alpha$ and the deviation angle $\sigma$ of the returning beam are related by $\alpha \approx \sigma/[2\cdot(n - 1)]$ for $\alpha < \SI{5}{\degree}$, where $n$ is the refractive index of the sample material.}
\label{fig:ac_double_pass}
\end{figure}

\begin{figure}[h!]
\centering
\includegraphics[scale=0.28]{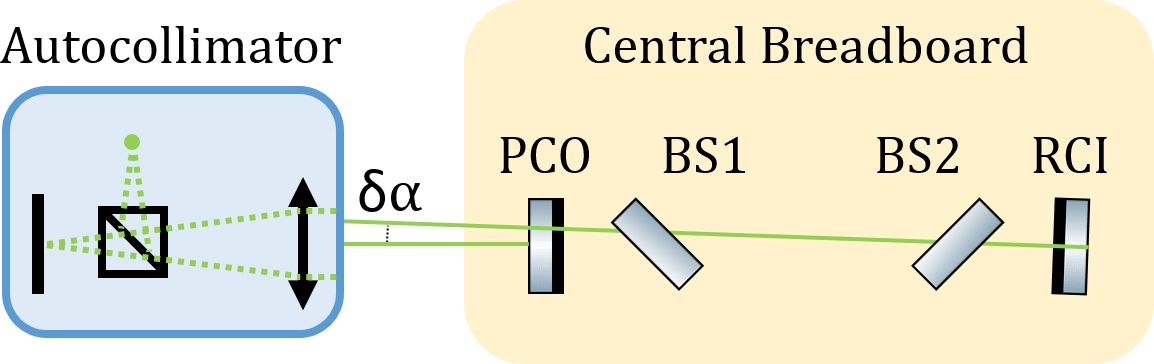}
\caption{Autocollimator measurement procedure for the ALPS II central optical bench (cf. Fig. \ref{fig:supra_fp}). In this illustrative examplary configuration, the planar mirror PCO and the two beamsplitters BS1 and BS2 have to be removed and the autocollimator registers the angle (zeroing) of the planar mirror RCI. PCO is then installed and properly tuned such that the autocollimator reading is within our angular specs. Finally, the plane-parallel BS1 and BS2 are installed at \SI{45}{\degree} incidence angle by means of dowel pins on the central optical bench. We note that due to the loose lateral specs, the \SI{45}{\degree} installation angle of the beamsplitters is not so critical. RCI is tilted in this illustration for the visibility of the angle $\delta \alpha$ between the two cavity mirror surfaces of interest, which in practice is to be minimized for our application.}
\label{fig:autocol_proc}
\end{figure}

An autocollimator is primarily used for angular measurements, based on which two mounting mechanisms have been developed and tested. The typical procedures for autocollimator measurements are illustrated in Figs. \ref{fig:ac_double_pass} and \ref{fig:autocol_proc}. Furthermore, we have come to develop two modified measurement procedures that can be performed in-situ (with the two planar mirrors always in place), the differential image analysis method and the split-aperture method.

\subsection{Differential image analysis method}
\begin{figure}[h!]
\centering
\includegraphics[scale=0.18]{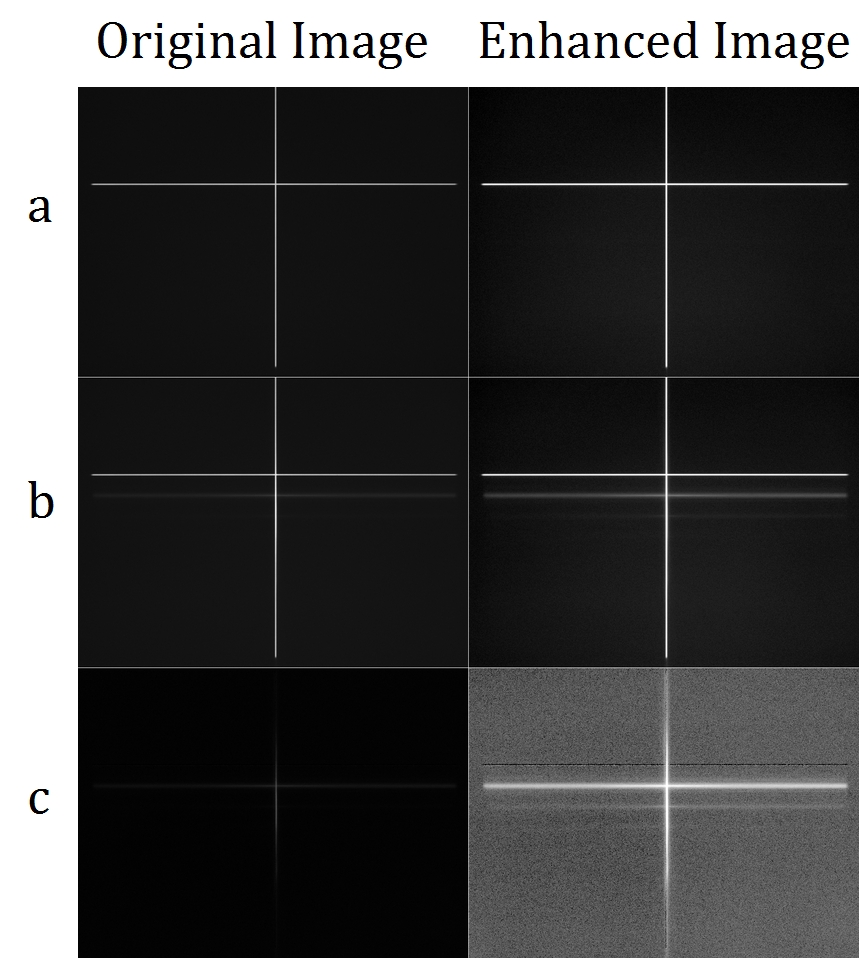}
\hspace{1cm}
\includegraphics[scale=0.45]{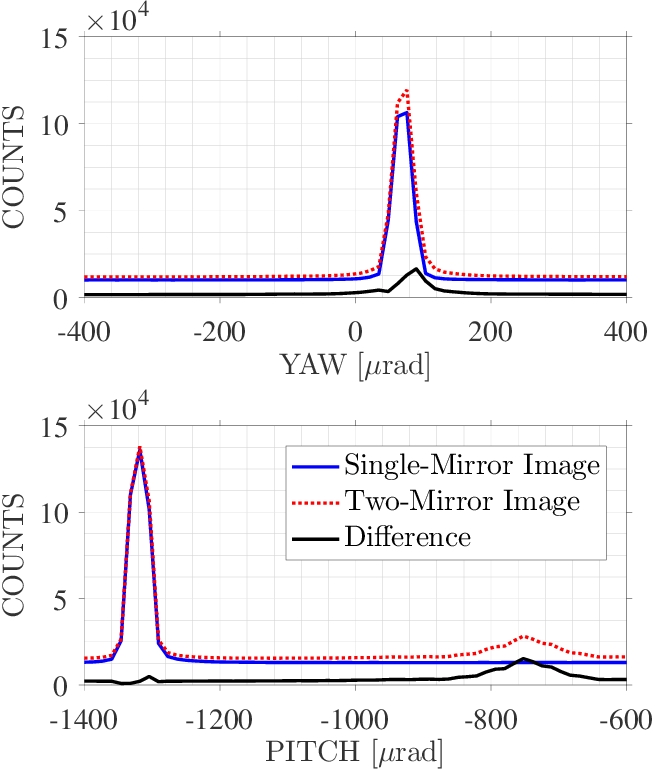}
\caption{Differential image analysis for the in-situ parallelism check with the autocollimator. \textit{(Left)} A visual comparison of the autocollimator crosshair images for (a) single-mirror, (b) two-mirror and (c) difference. Enhanced images are shown for the sack of visibility. \textit{(Right)} Cumulative analysis in yaw and pitch based on the original images. The hidden peak distribution in the two-mirror image is clearly visible after the subtraction.}
\label{fig:diff_image}
\end{figure}

The idea is to register two crosshair images with the autocollimator, one with only one cavity mirror and the other with two. By looking at the difference in the two images we derive the relative parallelism between the two mirrors in-situ. An exemplary demonstration is illustrated in Fig. \ref{fig:diff_image}. We note that the effectiveness of the differential image analysis method depends on the reflectivity of the coatings of the optics at the autocollimator wavelength. In our study the coatings are mostly only specified at the main laser wavelength of interest at \SI{1064}{\nano\meter}, while the autocollimator used green light. If devised properly, the differential image analysis method has the merit of providing a direct overall angular measurement of the central optical bench.

\subsection{Split-aperture method}
Taking advantage of the sufficiently large aperture of the autocollimator, it is possible to laterally dislocate the two cavity mirrors PCO and RCI on the central optical bench. This method is rather straightforward, however comes with the caveat of smaller clear aperture on the central optical bench, and possible measurement fallacies arising from mirror surface inhomogeneities which would require additional calibration effort. Partially limited by existing optics at hand, we choose the split-aperture method and have configured our test central optical bench accordingly for in-situ autocollimator measurements.

\subsection{Autocollimator}
An autocollimator (Trioptics TriAngle TA 300-57) at a wavelength $\lambda$ of \SI{525}{\nano\meter} is used. The autocollimator has a focal length of \SI{300}{\milli\meter}, an aperture diameter of \SI{50}{\milli\meter} and an image sensor with 
\SI[product-units = single]{780x580}{pixels} and $\SI[product-units = single]{8.2822 x 8.2817}{\micro\meter}$ pixel size. Using the autocollimator equation $d = 2 \alpha f$ where $\alpha$ is the angle, $f$ is the focal length and $d$ is the distance change on the image plane, the autocollimator used has a resolution of $\approx \SI{13.8}{\micro\radian/pixel}$. The Rayleigh criterion ($\lambda/D$, where $D$ is the aperture diameter) is $\approx \SI{10.5}{\micro\radian}$.

With sub-pixel image interpolation techniques, it is specified by the manufacturer to have a resolution of $\approx \SI{0.15}{\micro\radian}$, a repeatability of $\approx \SI{0.39}{\micro\radian}$ and an accuracy of $\approx \SI{3.64}{\micro\radian}$ \cite{TriAngle}.


\subsection{Rectangular optics with $\Pi$-shaped clamp}
\begin{figure}[h!]
\centering
\includegraphics[scale=0.3]{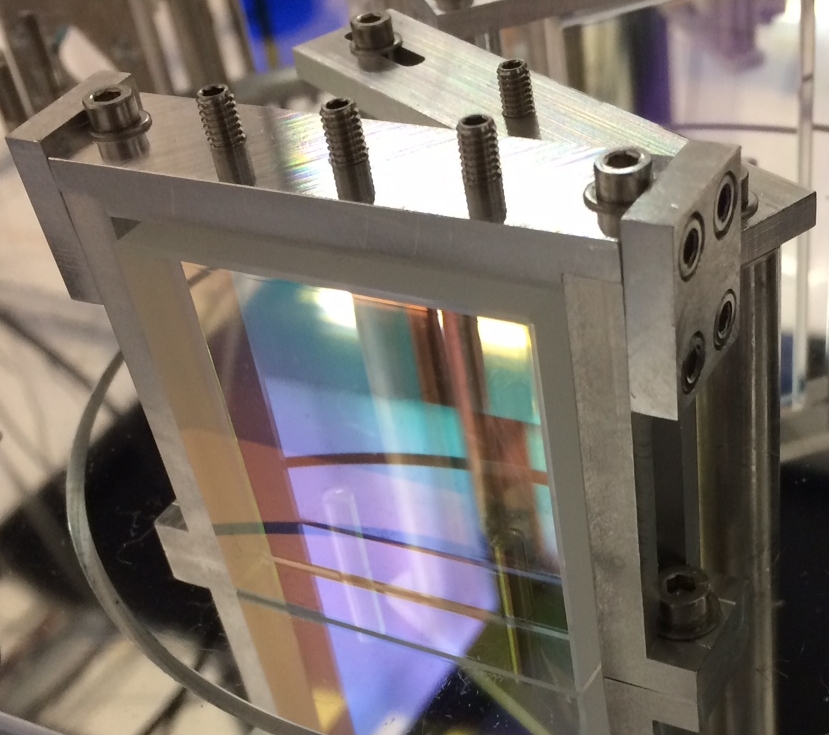}
\caption{$\Pi$-shaped clamp.}
\label{fig:pi_clamp}
\end{figure}

To achieve stringent parallelism and retain stability, an earlier concept during our study involves cuboid optics and a $\Pi$-shaped clamping frame structure. At least one adjescent surface of the cuboid is polished to $\SI{90}{\degree}\pm\SI{5}{\micro\radian}$  (right angle specification) with respect to the main optics surface. The $\Pi$-shaped clamping frame features spring-loaded ball-tipped screws to press the cuboid optics onto our test central optical bench, as illustrated in Fig. \ref{fig:pi_clamp}. The test central optical bench is custom made with smooth surface using ALPLAN (aluminum alloy 5083/AlMg4.5Mn0.7), and in conjunction with the cuboid optics automatically defines the parallelism in pitch. Dowel pins are then used to assist in defining the parallelsim in yaw.

The stability of the $\Pi$-clamped rectangular optics itself is tested to fulfill our angular specification. However, in practice, some complexity weighs in concerning the stringent right angle specification, which in the first place requires additional precision polishing and a thick mirror substrate. In our particular experiment it is observed that after the coating process the right angle specification met by the substrate is spoiled by the spillover of some coating material. As a workaround, aluminum foils have been used to shim the rectangular optics, but further complexity arises. First, the very hard material of the tipped coating spillover easily scratches the central optical bench. Secondly, shimming with aluminum foils seemingly leads to additional dependence on environmental factors such as humidity and poses stability issues.

\subsection{Introduction of a Thorlabs Polaris mirror mount and wedged optics}

Although properly coated rectangular optics that meet the right angle specifications should still serve our purpose, as an timely alternative in our study, the test central optical bench has been patched to allow for the introduction of a commercially available Thorlabs Polaris K1T mirror mount that is used to host a standard one-inch mirror acting as RCI. We note that, due to the nature of patching, this one-inch mirror is slightly wedged. Moreover, to allow for in-situ measurements with the split-aperture method, a smaller rectangular optics that does not meet the right angle specifications nor the angular specifications on plane-parallelism is used. Nevertheless, as we shall see, since we are able to characterize precisely the wedge angles and orientation, they have rather minimal impact on the purpose of our study. The beamsplitters, $\Pi$-clamped, on the other hand, are plane-parallel to within our angular specifications. 

The wedge angle of the optics on our test central optical bench, measured with the autocollimator as depicted in Fig. \ref{fig:ac_double_pass}, is summarized in Table \ref{table:wedge}.

\begin{table}[h!]
\centering
\begin{tabular}{ c | c | c | c | c }
  \hline			
  (\si{\micro\radian}) & PCO & BS1 & BS2 & RCI \\
  \hline
  Pitch	& -22 & 0.5 & 0.6 & 2 \\
  Yaw		& -3 & 0.6 & 1 & 46 \\
  \hline  
\end{tabular}
\caption{Wedge angle of the optics on the test central optical bench (Fig. \ref{fig:autocol_proc}), measured with the autocollimator. Yaw is defined as the rotation in the plane of the central optical bench.}
\label{table:wedge}
\end{table}

\subsection{Long-term stability of the central optical bench}
\begin{figure}[h!]
\centering
\includegraphics[scale=0.8]{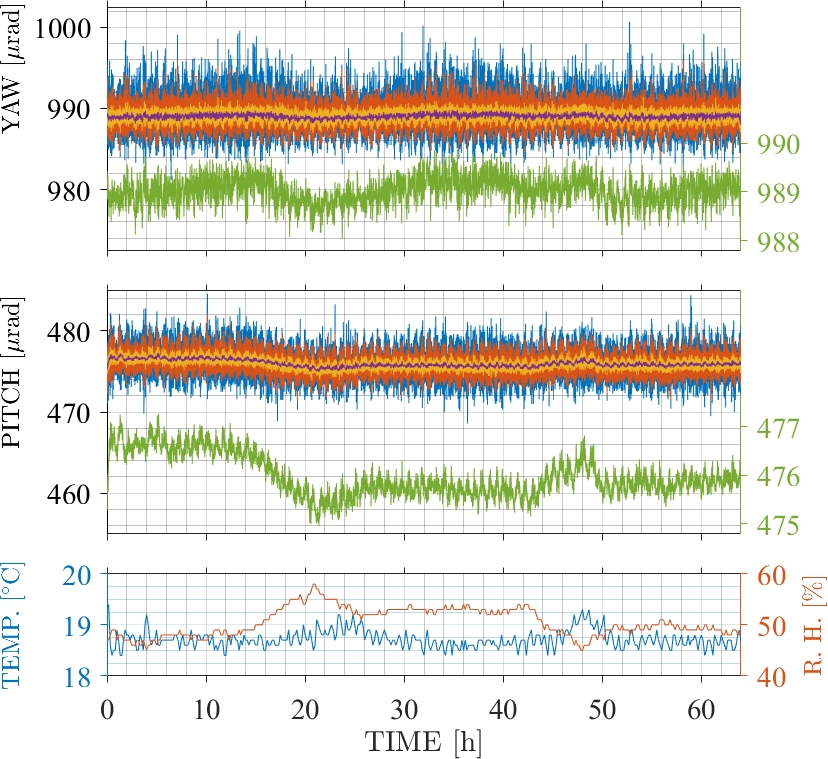}
\caption{Autocollimator measurement time series on the angle between a $\Pi$-clamped rectangular mirror and a standard one-inch mirror on a Thorlabs Polaris K1T mount. The most fluctuating yaw/pitch angle curves are raw data taken at $\approx \SI{110}{\milli\second}$ intervals. The smoother curves encompassed therein correspond sequentially to moving averages of 10, 100 and 1000 data samples. The curves with 1000 data samples are replotted (in green) using the enlarged y-axes on the right for better correlation visibility. The temperature and relative humidity (R.H.) data are sampled every \SI{10}{\minute}. There exists a strong correlation of $\num{-0.75}$ (Pearson correlation coefficient) between the pitch angle (averaged to \SI{10}{\minute} sampling period) and relative humidity, to be compared with the correlation of \num{-0.17} between the yaw angle and relative humidity. The correlation to temperature is \num{-0.05} for yaw and \num{-0.14} for pitch. The correlation calculations are not optimized with phase shifts.}
\label{fig:ta300-57_meas}
\end{figure}

Referring to Fig. \ref{fig:autocol_proc}, we configure the central optical bench such that:
\begin{itemize}
\item{PCO is a direct $\Pi$-clamped (i.e. without shimming) rectangular mirror and RCI is an one-inch mirror mounted to a Polaris K1T.}
\item{PCO and RCI are laterally sheared to allow for the split-aperture method.}
\item{PCO and RCI are intentionally angled to overcome the Rayligh criterion of th autocollimator.}
\end{itemize}
The autocollimator is then installed to continuously capture crosshair images. By doing so, we are able to characterize the stability of the whole setup and to give an estimate on the capability of the autocollimator in use. The time series results for more than 60 hours are shown in Fig. \ref{fig:ta300-57_meas}. Since these measurements are not performed in a well-protected lab environment (in contrast to ALPS II), we would like to direct the attention of the reader to the moving-averaged curves. As can be clearly seen from the green traces in Fig. \ref{fig:ta300-57_meas}, the test central optical bench meets our angular specification. The correlation to relative humidity is also clearly visible, based on which some caution might have to come into play in ALPS II regarding vacuum pump down effects.

\subsection{Capability of the autocollimator in use}
\begin{figure}[h!]
\centering
\includegraphics[scale=0.65]{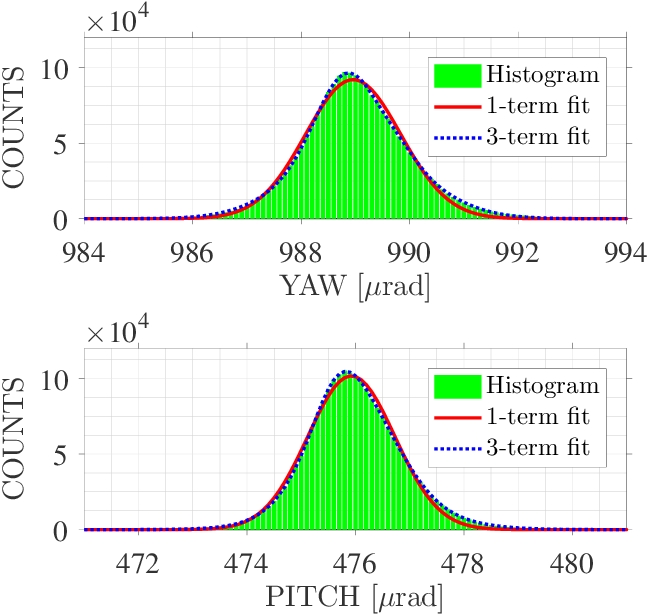}
\caption{Histogram and statistics of raw data in Fig. \ref{fig:ta300-57_meas}. The bin width is \SI{0.1}{\micro\radian}. The standard deviation $\sigma$ is \SI{1.03}{\micro\radian} for yaw and \SI{0.89}{\micro\radian} for pitch. The arbitrarily set mean values are not of interest. The different $\sigma$ values may be due to the difference in the number of rows (580) and the number of columns (780) on the image sensor. The $\sigma$ of pitch reduces slightly to \SI{0.84}{\micro\radian} if we remove its linear correlation of $\approx \SI{0.148}{\micro\radian/\percent}$ to relative humidity while that of yaw remains almost identical. The histograms are fit with Gaussian distributions ($a_i \cdot \textrm{exp}\{-[(x-b_i)/c_i]^2\}$) using 1 terms and 3 terms; the respective R-squared values are 0.9950 and 0.9999 for yaw, and 0.9962 and 1.0000 for pitch. The 1-term fit coefficients ($a_i, b_i, c_i$) are ($92112, 989, 1.241$) for yaw, and ($101751, 475.9, 1.135$) for pitch. The 3-term fit coefficients are ($47441, 988.8, 0.7629$), ($5947, 989.8, 0.4701$) and ($49808, 989, 1.584$) for yaw, and ($18922, 475.7, 0.565$), ($17219, 476.2, 1.704$) and ($71983, 475.9, 1.089$) for pitch. The goodness of the 3-term fits may suggest that there are three discrete creep states of the measurement setup, although caution should also be taken against overfitting.}
\label{fig:ta300-57_meas_stat}
\end{figure}

The same set of data may be used to derive an estimate on the capability of the autocollimator in use in the form of histogram, as shown in Fig. \ref{fig:ta300-57_meas_stat}. We see that the distribution of the raw data time series is rather normal, with a standard deviation of $\approx \SI{1}{\micro\radian}$ that may be compared to the manufacturer's specifications discussed earlier.

All in all, with sufficient averages, the autocollimator in use fulfills the measurement uncertainty requirements of ALPS II. We will next compare the autocollimator measurements to cavity-based measurements to derive an error estimate. Since cavity-based measurements are rather tedious, it is practical if they can be replaced with autocollimator measurements as far as our specifications are duly fulfilled. In the following we will compare these two independent measurements carefully.

\section{Cavity eigenmode spatial overlap}

Referring to Figs. \ref{fig:supra_fp} and \ref{fig:autocol_proc}, we have set up a downsized optics prototype with a cavity length of $\approx \SI{1}{\meter}$ while keeping the same dimension of the central optical bench, and we derive the spatial overlap of the cavity eigenmodes as follows:
\begin{enumerate}
\item{The input laser field (referred to as High-Power Laser in ALPS and in Fig. \ref{fig:supra_fp}) is frequency-stabilized to the fundamental Production Cavity Eigenmode (TEM\textsubscript{00}) with the Pound-Drever-Hall sensing technique \cite{PDH_ori, PDH_black}.}
\item{The beam transmitted by PCO passes through the components on the central optical bench on its way towards the RC.}
\item{The length of the RC is scanned, and the power in transmission is registered using a photodiode with high dynamic range electronic readouts.}
\end{enumerate}
The transmitted power versus length of the Regneration Cavity, known as a mode-scan, then carries the information of interest, which is simply the ratio between the transmitted $U_1$ power and the transmitted $U_0$ power in the eigenmode basis of the RC.

The downsized cavities have the following eigenmode parameters: $w'_{0,eig} \approx \SI{1}{\milli\meter}$, $\theta'_{0,eig} \approx \SI{0.33}{\milli\radian}$, and hence $z'_R \approx \SI{3}{\meter}$. The downsized cavities are therefore more discriminant for the lateral specifications but less discriminant for the angular specifications described in Eqn. \ref{eqn:ALPS_req}, following Eqn. \ref{eqn:U1}. In numbers, the power coupling from $U_0$ to $U_1$ due to angular misalignment is $\approx \num{2.6E-4}$ at the ALPS II angular specifications, which requires some high dynamic range optical power measurements.

\subsection{High dynamic range ($>$ 80dB) optical power measurements}

\begin{figure}[h!]
\centering
\includegraphics[scale=0.48]{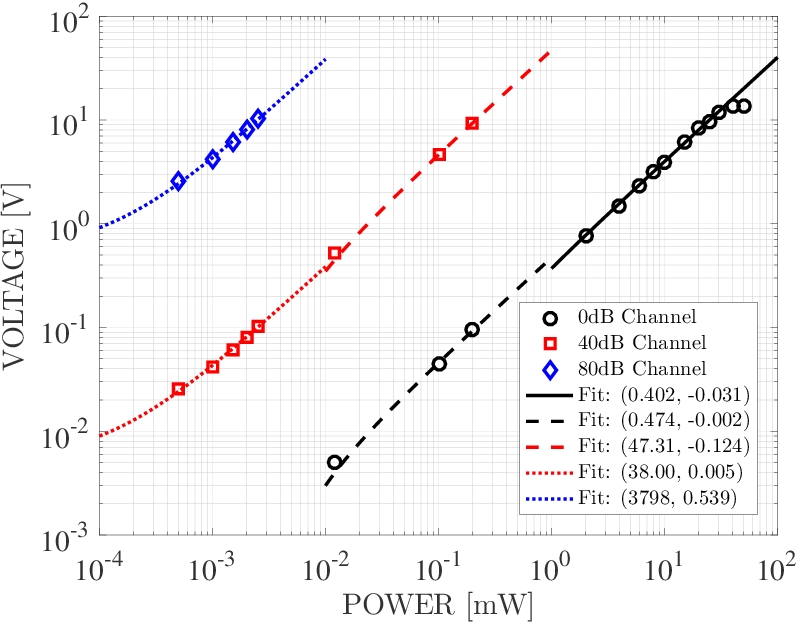}
\includegraphics[scale=0.48]{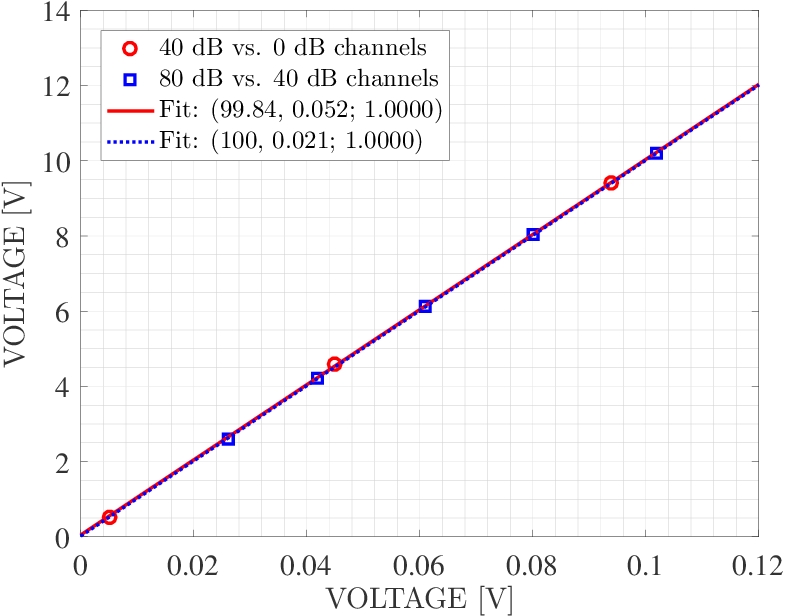}
\caption{Calibration measurements of the high dynamic range photodetector. \textit{(left)} Output voltage vs. input power. The nonlinear apperance at low input power level is likely due to stray light. Saturation of the electronics is visible toward high input power level. \textit{(right)} Gain of the 40 dB amplifiers. Since we are deriving relative measurements between the 0 dB, 40 dB and the 80 dB channels, it is the linearity of the amplifiers that is of more importance. Fig legends: (a, b; [R-squared]) with $y = ax + b$.}
\label{fig:hdr_pd}
\end{figure}

A photodiode with a transimpedance amplifier is used as our photodetector in transmission of the Regeneration Cavity. The voltage signal is further amplified by two 40 dB stages, resulting in three output channels: 0 dB, 40 dB and 80 dB. The calibration measurements are shown in Fig. \ref{fig:hdr_pd}.

\subsection{Spatial overlap optimization: wedge angle compensation}
As discussed earlier, PCO and RCI on our test central optical bench have wedge angles that are measured with the autocollimator to be out of our angular specifications. The high-reflectivity surface of RCI, which is on a adjustable mirror mount, therefore has to be intentionally tilted with respect to that of the PCO, following geometrical ray optics calculations, to maximize the overlap of the cavity eigenmodes. In other words, the planar cavity mirrors are no longer parallel to each other and therefore the cavities are not geometrically modematched. The rationale here is nevertheless to validate our tooling approach based on the use of the autocollimator, which in turn leads to geometrical modematching in a controlled and calibrated manner.

\begin{figure}[h!]
\centering
\includegraphics[scale=0.7]{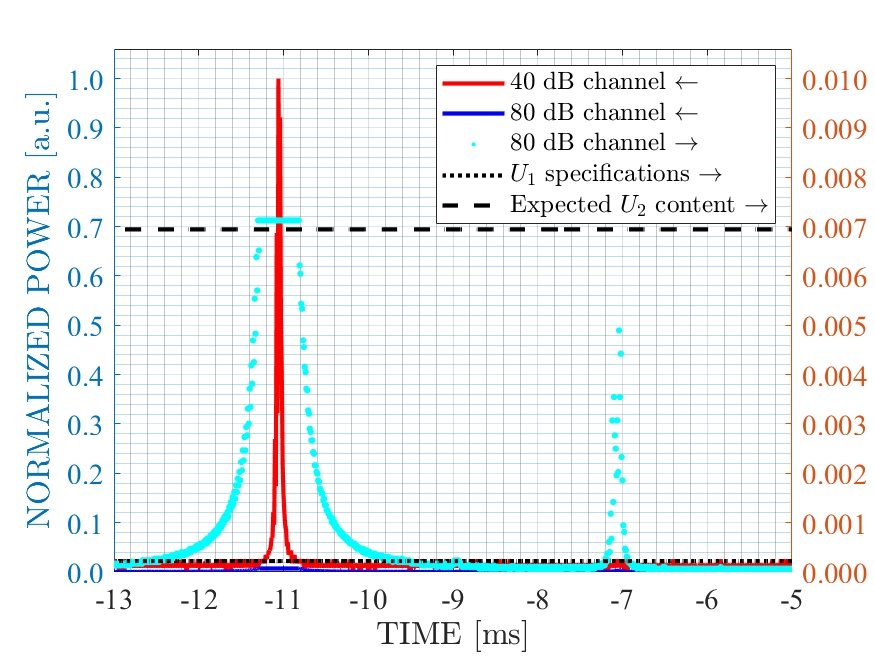}
\includegraphics[scale=0.7]{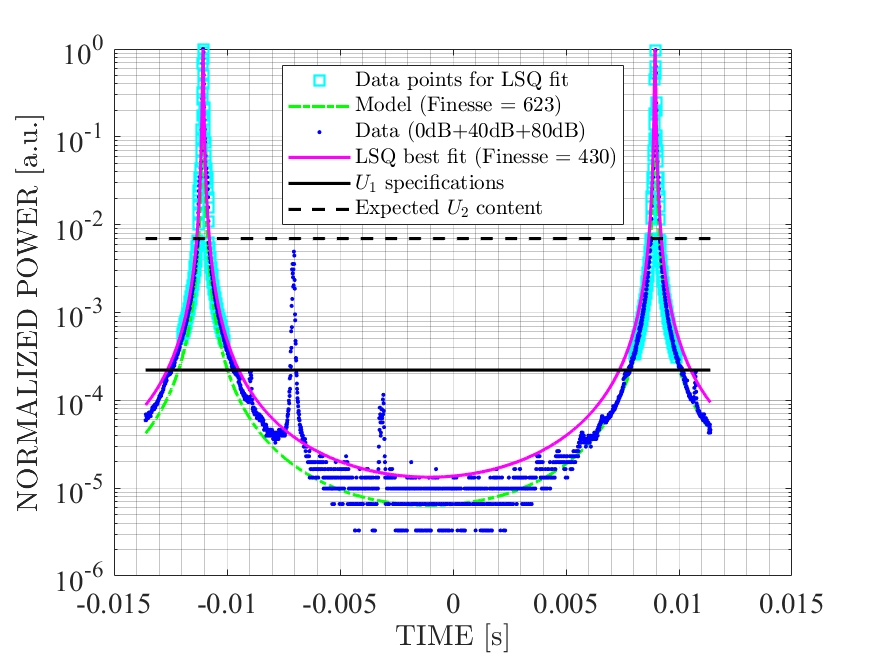}
\caption{Eigenmode spatial overlap after fine-tuning. The higher-order-mode spacing corresponds to $\approx \SI{2}{\milli\second}$ in the scan. The ($U_0$, $U_1$, $U_4$) transmissions are at $\approx$(\SI{-11}{\milli\second}, \SI{-9}{\milli\second}, \SI{-7}{\milli\second}) in the scan. An order-4 transmission is also visible at $\approx$\SI{-3}{\milli\second}. \textit{(top)} Raw data with zoom-in for clarity. The direction of the arrow in the legend indicate the y-axis to be read. \textit{(bottom)} The complete (TEM\textsubscript{00} to TEM\textsubscript{00}) modescan with fit analysis. The modelled Finesse value does not consider losses. LSQ: least-squares.}
\label{fig:overlap_result}
\end{figure}

\begin{figure}[h!]
\centering
\includegraphics[scale=0.7]{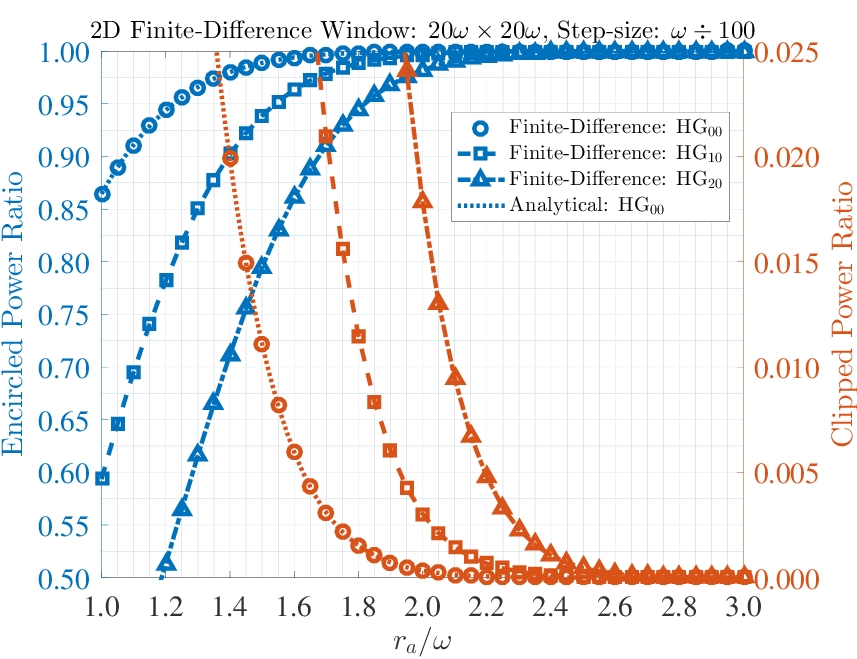}
\caption{Differential clipping loss of high-order-modes imposed by finite circular aperture. $r_a$: radius of the aperture; $\omega$ laser beam spot size on the aperture.}
\label{fig:hom_clipping}
\end{figure}

\subsection{Spatial overlap optimization: the lateral degree of freedom}

To obtain a correct account on the angular misalignment, the effect of lateral misalignment has to be minimized, which rely on the fine-tuning of the two cavity curved mirrors (see Fig. \ref{fig:supra_fp}). The PZT mounts of the curved mirrors have three pivot points in a equilateral triangule constellation. A matrix amplifier is used to convert actuation input signals for piston movement, yaw rotation and pitch rotation to correponding actuations on the three pivot points. The autocollimator is used to calibrate the matrix amplifier; with the calibrated matrix amplifier we assure a pure piston motion of RCO during the modescans. The PZT mounts sit on a high-load pitch and yaw platform (Thorlabs PY004/M). The platform is used for rough alignments along with some mechanical alignment aid to arrive at rough spatial overlap, after which precise beam-walking is performed using the PZT mount of RCO to minimize the transmitted $U_1$ power. 

\subsection{Spatial overlap optimization: results}

The resultant eigenmode spatial overlap measurement is shown in Fig. \ref{fig:overlap_result}. Although marginal, we see that the measured $U_1$ content indicates the fulfillment of our specifications. We see also that the measured $U_2$ transmission is $\approx \SI{30}{\percent}$ lower than expected, derived from $z'_{R} \approx \SI{3}{\meter}$ and $\delta z_{0,eig} \approx \SI{0.5}{\meter}$ with the use of Eqn. \ref{eqn:U2}. This may be partially, if not best, explained by the differential clipping loss of high-order-modes imposed by the finite aperture of the photodetector, as well as the finite aperture of the overall cavity setup. Some two-dimensional finite-difference calculations are made, as shown in Fig. \ref{fig:hom_clipping}.


\section{The QPD assembly}

After the fulfillment of the angular specifications, we now turn our focus to the lateral specifications. Once optimal cavity eigenmode spatial overlap is obtained, the next operational step in ALPS II is to register and lock the lateral positions of the beam on the QPDs (see Fig. \ref{fig:supra_fp}). The wall is then inserted between the two cavities to start the science data taking of ALPS II. When in doubt one can always temporarily remove the wall (which in practice is a light-tight shutter) to check the spatial overlap and to re-optimize it whenever needed. Nevertheless, the more stable the setup intrinsically is, the higher the duty cycle is, and the more efficient the experiment is.

It is therefore beneficial to have stable QPD assemblies despite the much less stringent requirements in the lateral degree of freedom of ALPS II. Another important aspect of the QPD assembly is scattering. At targeted search sensitivity of \SI{2E-11}{GeV^{-1}}, the experimental design of ALPS II anticipates an extremely low photon flux of $\approx$ 2 photons per day. It is therefore natural to minimize scattering with best effort from the source, alongside dedicated scattering absorbers and attenuators that prevent scattered photons from reaching the single-photon counter.

\begin{figure}[h!]
\centering
\includegraphics[scale=0.3]{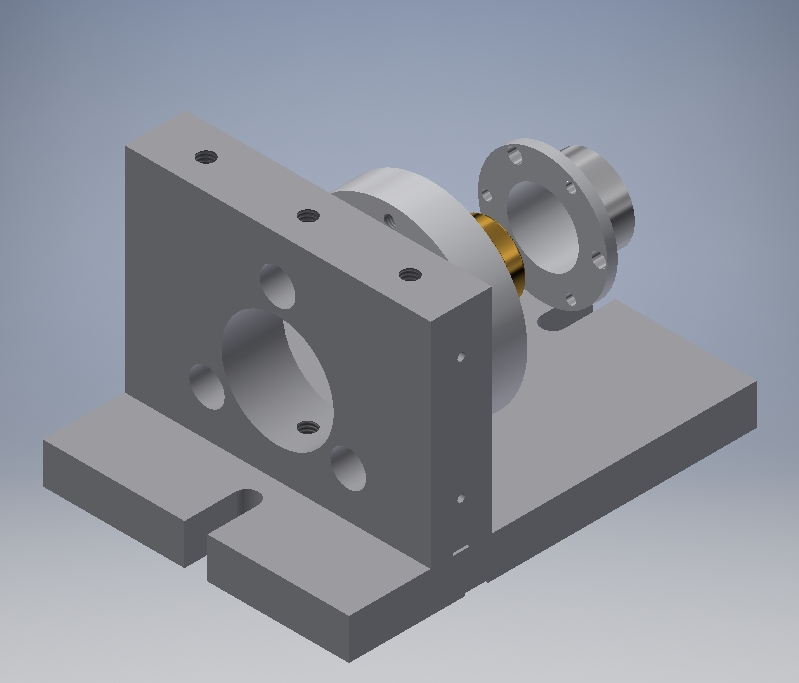}
\includegraphics[scale=0.3]{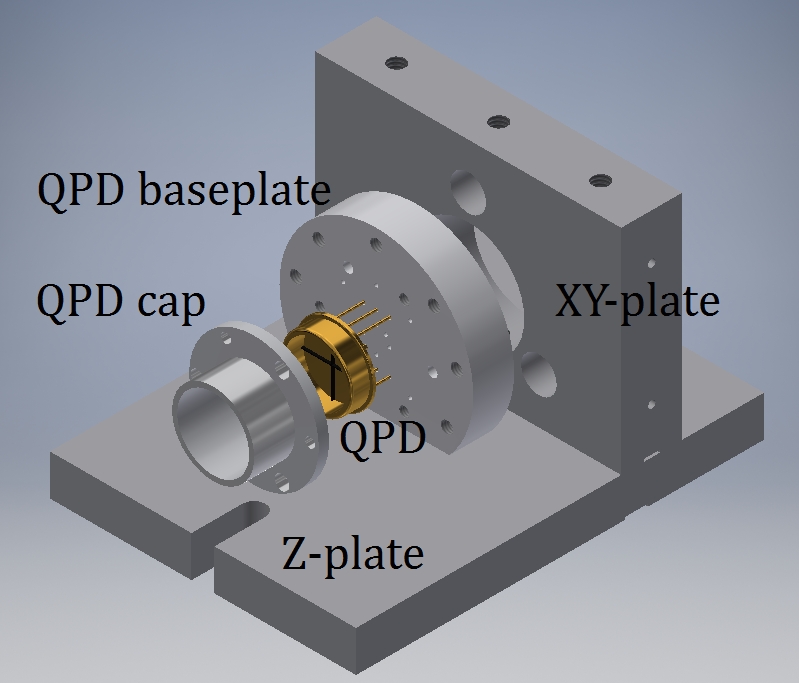}
\caption{The QPD assembly. The QPD (in TO-8 package shown here) is sandwiched between a baseplate and a cap with three M2.5 screws. The sandwich is then fixed to the XY-plate with three M3 screws through clearance holes with loose tolerance (\SI{7}{\milli\meter} diameter). The sandwich can take QPD packages with a diameter up to $\approx \SI{25.4}{\milli\meter}$ if set screw fixation of the QPD from the side is also considered. The loose holes allow for some tuning in the XY plane. The XY-plate is fixed to the Z-plate, and the Z-plate in turn on the central optical bench, all with socket cap M4 screws. In this study a Z-plate with different mounting hole configuration to the central optical bench is used to adapt to our test central optical bench. The QPD baseplate is made of material with Galvanic isolation. For ALPS II PEEK (polyether ether ketone) will be used for vaccum compatibility, while POM (polyoxymethylene) is used as the prototype for this study. We note that alumina has better heat conductibility if required. The other parts are made of aluminum alloy.}
\label{fig:qpd_mount}
\end{figure}

In order to be able to center the quadrant photo-detector to the laser beam, thus maximizing the linear readout range and minimizing scattering due to beam clipping, and to retain the required stability, we adopt a mounting design that is similar to the power stabilization photodiode box used in Advanced Virgo \cite{AdV_2015, Merzugui+Cleva_2017}. The mount is re-designed to be compatible with large-diameter QPDs and can be fine-adjusted along all three axes X, Y, and Z before being firmly fixed. A CAD rendering is shown in Fig. \ref{fig:qpd_mount}. Non-normal incidence angle together with dedicated beam dump for specular reflection can also easily be arranged with this design.

\begin{figure}[h!]
\centering
\includegraphics[scale=0.6]{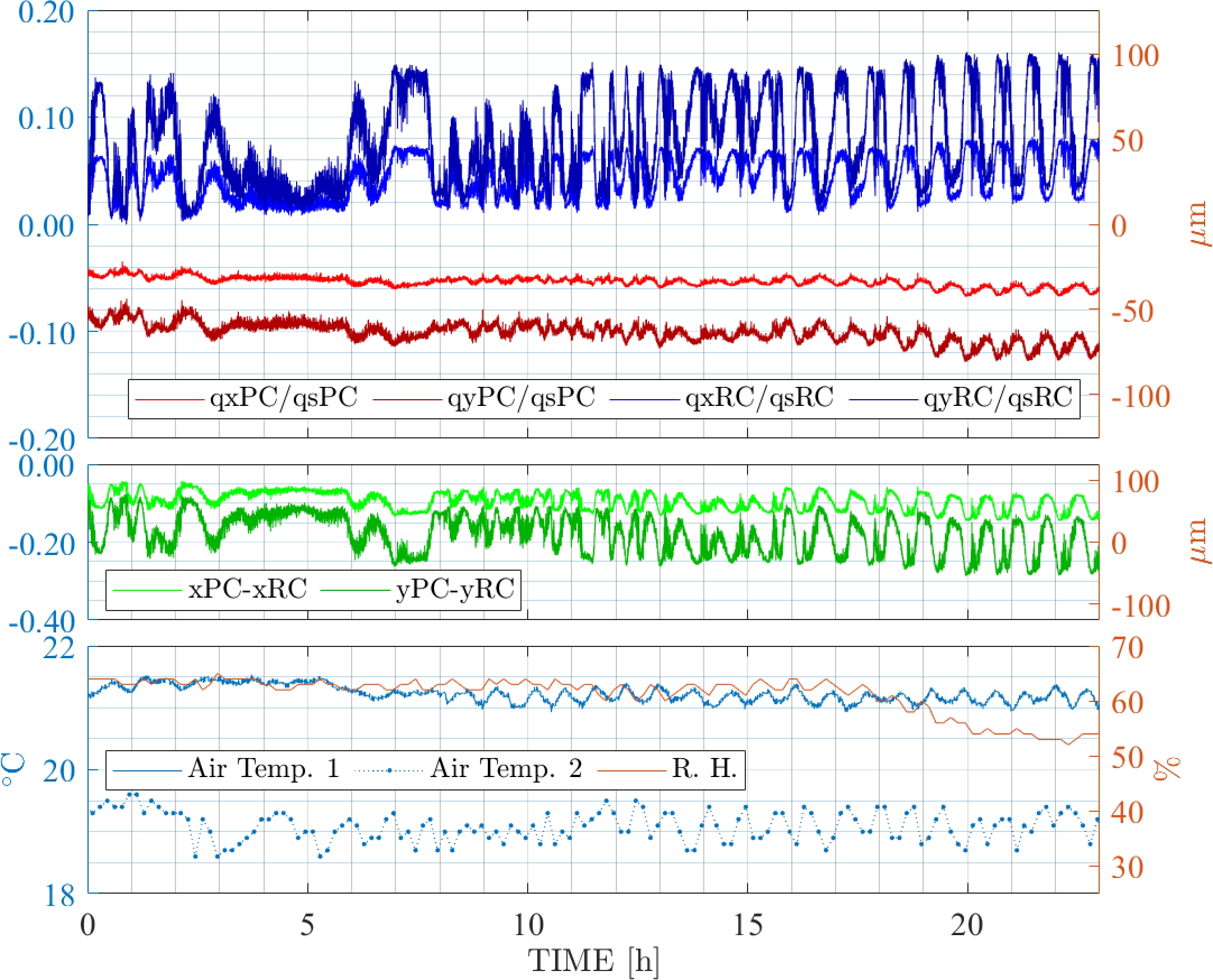}
\caption{QPD readings with Production Cavity installed. qx: quadrant difference in the x-direction; qy: quadrant difference in the y-direction; qs: quadrant sum; PC: Production Cavity (pick-off beam of BS1); RC: Regeneration Cavity (pick-off beam of BS2); xPC = qxPC/qs/PC, likewise for yPC, xRC and yRC. R. H.: relative humidity. The y-axes in \si{\micro\meter} unit show calibrated positional values of the other corresponding y-axes, with arbitrary units and zeroed in the middle panel.}
\label{fig:qpd_stability_PC}
\end{figure}

\begin{figure}[h!]
\centering
\includegraphics[scale=0.6]{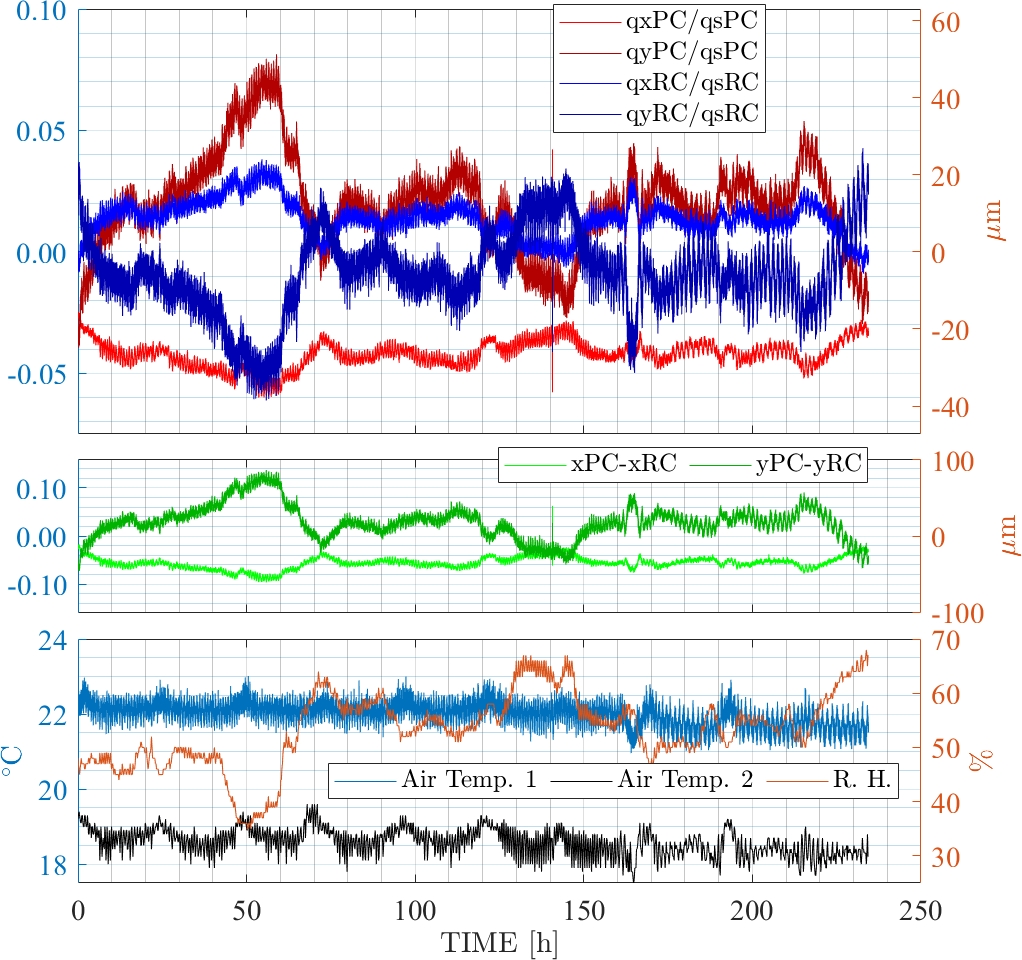}
\caption{QPD readings without Production Cavity. See Fig. \ref{fig:qpd_stability_PC} for explanation on legend.}
\label{fig:qpd_stability}
\end{figure}

Two QPD assemblies are installed onto the test central optical bench to sample the beam picked off by BS1 and BS2 (Fig. \ref{fig:autocol_proc}), with a laser beam being injected from the PCO side. The stability of the setup is then estimated by comparing the reading of the two QPDs. Two sets of measurements have been performed, one with the Production Cavity installed and the other without, as shown in Figs. \ref{fig:qpd_stability_PC} and \ref{fig:qpd_stability}. Although more delicate analysis such as correlation may be performed on these data, the main message should already be clear that even in a not-so-controlled lab environment the difference in the reading of the two QPDs are of the order of \SI{100}{\micro\meter}, which is well below our lateral specifications in Eqn. \ref{eqn:ALPS_req}. We see that the difference in the readings are driven by air temperature and relative humidity, while we should note that in the case of the cavity-less measurement, relative angular tilt between the laser and the central optical bench also results in a difference in the QPD readings.

\section{Conclusions and outlook}
To achieve the agressive optical gain anticipated in ALPS II, the two Fabry-Perot cavities on each side of the wall have to remain in strict resonance and their fundamental eigenmodes have to overlap maximally in space in the absence of the optics that are typically inserted in between for modematching. In this study we show that the latter ALPS II requirement, which we denote as the geometrical modematching of optical cavities, is met in our test setup.

The proposed geometrical modematching concept is operationally straightforward and modularized: strict plane-parallel mirrors are stably mounted parallel to govern the angular degree of freedom of the cavity eigenmodes while the lateral degree of freedom is controlled using dedicated quadrant photodetectors. Moreover, instead of tedious cavity-based metrology, we show that an autocollimator may be used to largely streamline the installation procedure. The proposed concept and demonstrated tooling approach are therefore to our belief fully compatible with the operation of ALPS II.

Several lessons are learnt during the study. Most notably, we see that when stringent angular requirement of few microradians is to be considered, care is to be taken in all aspects in the preparation of the central optical bench, from the mirror coating process, optics handling and component cleanliness, to environmental fluctuations such as temperature and humidity. The same applies to its characterization that, for example, the subtle differential beam clipping loss of eigenmodes of various orders is to be born in mind, and the electronics as well as the actuators are to be free from offsets and calibrated properly.

The presented study provides a clear approach for the geometrical modematching of optical cavities which is particularly useful for ALPS II and future cavity-enhanced light-shining-through-a-wall experiments. Additional features are also to be implemented on the central optical bench to achieve the resonance condition between the two Fabry-Perot cavities in ALPS II, allowing for the detection scheme with either a transition-edge sensor \cite{Baehre+2013} or heterodyne interferometry \cite{PhysRevD.99.022001}. That however only concerns the optical design, which may be fully supported by the presented optomechanics and does not interfere with our geometrical modematching concept.

\section*{Funding}
This work was funded by the Deutsche Forschungsgemeinschaft (DFG, German Research Foundation).

\clearpage



\bibliography{osa_overlap}






\end{document}